\documentclass[reprint,prl,aps,superscriptaddress,twocolumn,10pt]{revtex4-1}
\usepackage{amsmath}
\usepackage[english]{babel}
\usepackage{graphicx}
\usepackage{color}
\usepackage[dvipsnames]{xcolor}
\usepackage{sidecap}

\usepackage[colorlinks,
            citecolor = blue, 
						linkcolor = blue,
						urlcolor  = blue,
						anchorcolor = blue,
						breaklinks = true
						]{hyperref}

\usepackage{soul}
\usepackage{ amssymb }

\newif\ifdraft
\drafttrue
\def \ETH{Institute for Quantum Electronics, ETH Z\"urich, CH-8093 Z\"urich, Switzerland}
\def \NIMSRCFM{Research Center for Functional Materials, National Institute for Materials Science, Tsukuba, Ibaraki 305-0044, Japan}
\def \NIMSICMN{International Center for Materials Nanoarchitectonics, National Institute for Materials Science, Tsukuba, Ibaraki 305-0044, Japan}

\begin{document}

% ----- Title -----

\title{Spin-valley relaxation dynamics of Landau-quantized electrons in MoSe$_2$ monolayer} 

\author{T. Smole\'nski}
\affiliation{\ETH}

\author{K.~Watanabe}
\affiliation{\NIMSRCFM}

\author{T.~Taniguchi}
\affiliation{\NIMSICMN}

\author{M. Kroner}
\affiliation{\ETH}

\author{A. Imamo\u{g}lu}
\affiliation{\ETH}

\begin{abstract}
Non-equilibrium dynamics of strongly correlated systems constitutes a fascinating problem of condensed matter physics with many open questions. Here we investigate the relaxation dynamics of Landau-quantized electron system into spin-valley polarized ground state in a gate-tunable MoSe$_2$ monolayer subjected to a strong magnetic field. The system is driven out of equilibrium with optically injected excitons that depolarize the electron spins and the subsequent electron spin-valley relaxation is probed in time-resolved experiments. We demonstrate that the relaxation rate at millikelvin temperatures sensitively depends on the Landau level filling factor: it becomes faster whenever the electrons form an integer quantum Hall liquid and slows down appreciably at non-integer fillings. Our findings evidence that valley relaxation dynamics may be used as a tool to investigate the interplay between the effects of disorder and strong interactions in the electronic ground state. 
\end{abstract}

\maketitle

Over the last decade, there has been an explosive growth of research investigating two-dimensional (2D) semiconductors such as transition metal dichalcogenide (TMD) monolayers and their van der Waals heterostructures~\cite{Mak_PRL_2010,Xu2014}. This system features unique optical properties owing to ultralarge exciton binding energy~\cite{Berkelbach_PRB_2013,Chernikov_PRL_2014} as well as the existence of valley pseudospin degree of freedom~\cite{Xiao_PRL_2012,Cao_NatNano_2012,Zeng_NatNano_2012,Mak_NatNano_2012} that is locked to the spin by a strong spin-orbit coupling. In parallel, TMD heterostructures offer a fertile ground for investigations of correlated electronic states that arise due to strong Coulomb interactions. This has been recently demonstrated by several breakthrough experiments~\cite{Tang2020,Regan2020,Shimazaki2020,Wang2020,Shimazaki_PRX_2021} evidencing the formation of Mott-like correlated insulating (CI) states in twisted TMD hetero- and homo-bilayers. Unlike magic-angle twisted bilayer graphene (MATBG)~\cite{Cao_Nature_2018_1,Cao_Nature_2018_2,Lu_Nature_2019}, even fractional fillings of the TMD moir\'e superlattices show a CI behavior, providing a direct evidence for the dominant role played by long-range interactions that break discrete translation symmetry~\cite{Regan2020,Xu_Nature_2020}. Remarkably, even in TMD monolayers the Coulomb interactions between the itinerant electrons at densities not exceeding a few $10^{11}~\mathrm{cm}^{-2}$ turn out to be strong enough to allow the electrons to spontaneously break continuous translational symmetry and form a Wigner crystal (WC), as recently discovered by Refs~\cite{Smolenski_arXiv_2020,Zhou_arXiv_2020}. In parallel, the formation of fractional quantum Hall states in a TMD monolayer has also been evidenced under the influence of a strong magnetic field~\cite{Shi_NatNano_2020}.

Despite a rapid progress in exploration of strong electronic correlations in MATBG and TMD heterostructures, the prior research focused primarily on the ground-state properties. Many of the interesting open questions in condensed matter physics, however, concern the non-equilibrium dynamics of strongly correlated systems. In the context of TMD systems, a key question that would determine the utility of the valley degree-of-freedom is the relaxation dynamics of an electron or hole system following an inter-valley excitation. For WSe$_2$/WS$_2$ hetero-bilayer, such a hole spin-valley relaxation was shown to slow down upon the formation of a CI state at an integer filling of the moir\'e superlattice~\cite{Regan2020}. In case of TMD monolayers, even though the spin-valley relaxation dynamics has been investigated in several prior experiments~\cite{Crooker_NPhys_2015,Hsu_NCommun_2015,Yan_PRB_2017,McCormick_2017,Song_NL_2016,Dey_PRL_2017,Goryca_SciAdv_2019,Goryca_PRM_2021}, the effects of ground-state electronic correlations on this relaxation remained elusive.

Here we study the temporal dynamics of an excited state of Landau-quantized electron system in a charge-tunable MoSe$_2$ monolayer under high external magnetic fields ($B=14$~T). This state is prepared optically by means of a resonant injection of excitons that interact with itinerant electrons and lead to sizable depolarization of their spins. Our time-resolved pump-probe experiments reveal that the electronic spin relaxation rate exhibits striking, periodic oscillations with the Landau level (LL) filling factor $\nu$ at mK temperatures. The fast relaxation for integer quantum Hall (IQH) states together with its striking slowing down for a WC at $\nu\lesssim0.5$~\cite{Smolenski_arXiv_2020} suggest that $\nu$-dependent correlations in the electronic ground state may be responsible for the observed effects.

\begin{figure*}[t]
	\includegraphics{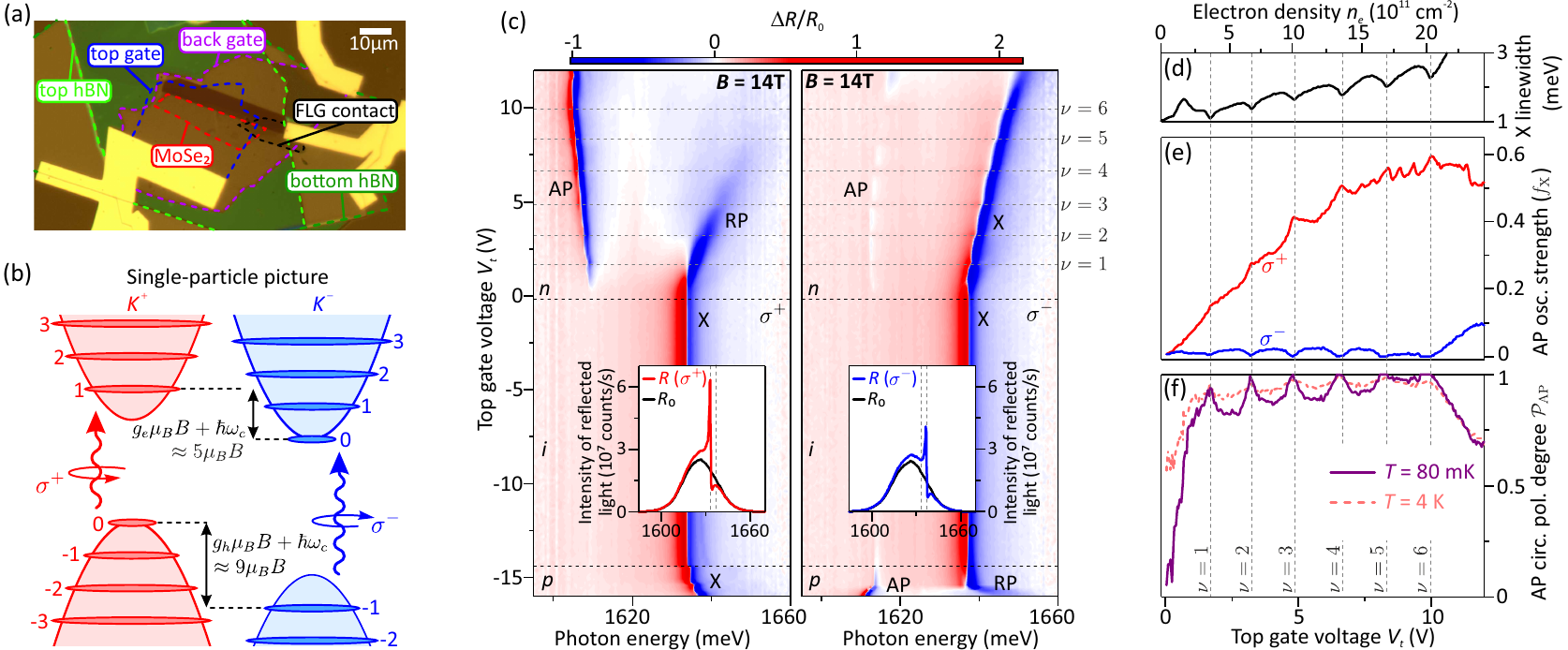}
	\caption{(a)~Optical micrograph of the investigated device that consists of a dual-graphene-gated MoSe$_2$ monolayer encapsulated between two hBN layers (dashed lines mark the boundaries of subsequent flakes). (b)~Schematic illustrating the lowest-energy spin-orbit-split bands in a MoSe$_2$ monolayer at a finite magnetic field in a single-particle approximation (i.e., in the absence of many-body interactions). The energy splitting $g^*_{e,h}\mu_BB$ between the lowest electron (hole) LLs in the $K^+$ and $K^-$ valleys is given by the sum of the cyclotron energy $\hbar\omega_c=\hbar eB/m_{e,h}^*$ and the electron (hole) Zeeman term, corresponding to effective $g$-factor of $g_e^*\approx5$ ($g_h^*\approx9$)~\cite{Xu2014,Aivazian_NP_2015,MacNeil_PRL_2015,Koperski_2DM_2018} for the electron and hole effective masses $m_e^*\approx0.7m_0$~\cite{Larentis2018} and $m_h^*\approx0.6m_0$~\cite{Zhang2014,Goryca_NC_2019}. (c)~Color-scale maps showing the top-gate-voltage evolution of the white-light reflectance contrast spectra $\Delta R/R_0=(R-R_0)/R_0$ measured at $T=80$~mK, $B=14$~T in the two circular polarizations:~$\sigma^+$~(left) and~$\sigma^-$~(right). Horizontal dashed lines mark the onsets of doping of conduction and valence bands with carriers as well as subsequent integer filling factors on the electron-doped side. The insets in both maps depict example reflectance spectra $R$ and $R_0$ acquired, respectively, on and off the MoSe$_2$ flake at $V_t=-1$~V (when the MoSe$_2$ is devoid of itinerant carriers), based on which the reflectance contrast is evaluated (vertical dashed lines mark the exciton energies in both polarizations). \mbox{(d-f)}~Electron-density-dependence of the linewidth of $\sigma^-$-polarized X resonance~(d), oscillator strengths $f_\mathrm{AP_\pm}$ of the $\sigma^\pm$-polarized AP resonances~(e), and the circular polarization degree $\mathcal{P}_\mathrm{AP}$ of the AP~(f) determined based on fitting the lineshapes of the transitions in panel~(c) following the procedures described in SM~\cite{SM}. The AP oscillator strengths are expressed relative to the oscillator strength $f_\mathrm{X}$ of the X resonance in the charged-neutral regime. The dashed line in~(f) presents the AP polarization degree determined at the same spot and magnetic field, but at an elevated temperature of $T=4$~K.\label{fig:Fig1}}
\end{figure*}

The analyzed device consists of a charge-tunable MoSe$_2$ monolayer that is encapsulated between two hBN layers and two few-layer-graphene flakes serving as top and back gate electrodes (see Fig.~\ref{fig:Fig1}(a) and Ref.~\cite{Smolenski_arXiv_2020} for details). For the experiments, the device was mounted in a dilution refrigerator with a monomode-fiber-based optical access allowing to perform polarization-resolved, magneto-optical experiments at a base temperature of either 80~mK or 4~K~\cite{Smolenski_arXiv_2020}. In case of the reflectance measurements, the sample was illuminated with a light emitting diode (LED) featuring a center wavelength of 760~nm and 20-nm linewidth. The resonant fluorescence (RF) and photoluminescence excitation (PLE) experiments were in turn performed with the use of a single-frequency, continuous-wave (CW) Ti-sapphire that was spectrally-broadened using an electro-optic phase modulator with a $\sim$20~GHz drive (to reduce the coherence length and the related etaloning, while retaining narrow linewidth $<0.1$~meV). All of the results presented in the main text were obtained at $B=14$~T (see Supplemental Material (SM)~\cite{SM} for complementary datasets acquired on a different device).

Fig.~\ref{fig:Fig1}(c) shows a representative top-gate-voltage ($V_t$) evolution of the circular-polarization-resolved reflectance contrast $\Delta R/R_0$ spectra taken at $T=80$~mK. In the charge-neutral regime (at $-14\ \mathrm{V}\lesssim V_t\lesssim0\ \mathrm{V}$), the spectra display a single, bare exciton resonance (X) that is split between the two circular polarizations by $g_X^*\mu_BB$ due to the valley-Zeeman effect with $g^*_X\approx4.3$~\cite{Li_PRL_2014,Aivazian_NP_2015,Srivastava_NP_2015}. Similarly, the valley degeneracy of both conduction and valence bands is lifted for $B \neq 0$. In a single-particle approximation [see Fig.~\ref{fig:Fig1}(b)], the resulting splitting $g^*_{e,h}\mu_BB$ of the lowest electron (hole) LLs is $\approx4$~meV ($\approx7$~meV) at $B=14$~T, assuming an effective $g$-factor of $g_e^*\approx5$ ($g_h^*\approx9$)~\cite{Xu2014,Aivazian_NP_2015,MacNeil_PRL_2015,Koperski_2DM_2018}. Therefore, the spin-valley splitting exceeds the thermal energy $k_BT$ by more than an order of magnitude even at $T=4$~K. Consequently, at low doping densities the itinerant electrons (holes) are expected to be fully spin-polarized and fill the states in $K^-$ ($K^+$) valley for $B>0$. Under such conditions only the excitons in the opposite $K^+$ ($K^-$) valley can get dressed into attractive (AP) and repulsive (RP) Fermi polarons~\cite{Sidler2017,Efimkin2017}, leading to the emergence of a red-shifted AP resonance exclusively in $\sigma^+$ ($\sigma^-$) polarization. 

While the above picture remains in perfect agreement with the optical response measured on the hole side (at $V_t\lesssim-14$~V), in case of the electron-doping (at $V_t\gtrsim0$~V) we clearly observe the AP resonances in both polarizations. Although the $\sigma^+$-polarized resonance (AP$_+$) is much stronger than its $\sigma^-$-polarized counterpart (AP$_-$), the latter exhibits pronounced intensity oscillations as the electron density $n_e$ is varied, indicating that spin-valley polarization depends on the LL filling factor $\nu$. To quantitatively analyze this effect, we fit the lineshapes of both resonances with a transfer-matrix approach (see SM~\cite{SM} for details), which allows us to extract their oscillator strengths $f_\mathrm{AP_\pm}$ being directly proportional to the densities $n_e^{\mp}$ of electrons residing in $K^\mp$ valleys~\cite{Sidler2017,Glazov_JCP_2020,Imamoglu_CRP_2021}. Figs~\ref{fig:Fig1}(e,f) display gate-voltage dependencies of the determined $f_\mathrm{AP_\pm}$ along with the corresponding polarization degree $\mathcal{P}_\mathrm{AP}=(f_\mathrm{AP_+}-f_\mathrm{AP_-})/(f_\mathrm{AP_+}+f_\mathrm{AP_-})$.  The AP intensity oscillations are directly correlated with the LL filling factor, as revealed by their coincidence with Shubnikov-de Haas oscillations of the $K^-$ exciton linewidth [Fig.~\ref{fig:Fig1}(d)] \footnote{We use the minima of the $K^-$ exciton linewidth, which occur precisely at integer $\nu$~\cite{Smolenski2019}, to calibrate the electron density $n_e$, see SM~\cite{SM} for details.}. Specifically, the AP$_-$ resonance is stronger around half-integer $\nu$, and becomes barely discernible for integer~$\nu$ (until $\nu=6$, beyond which the Fermi energy exceeds the effective valley-Zeeman splitting of the conduction band). These changes coincide with the periodic variations of the slope of the AP$_+$ intensity increase, demonstrating that the electrons become partially spin-valley depolarized each time the highest-energy LL is partially occupied. Such depolarization turns out to be particularly prominent at $\nu<1$, where $\mathcal{P}_\mathrm{AP}$ steeply decreases for lower $n_e$, reaching almost zero in the zero-density limit. Interestingly, this initial drop of the polarization degree becomes suppressed upon rising the temperature to $T=4$~K. This observation is in stark contrast to naive expectation that electronic spin depolarization would be enhanced by thermal fluctuations. At elevated $T$ the density-dependent polarization variation also becomes clearly less pronounced [see Fig.~\ref{fig:Fig1}(f)].

\begin{figure}[t!]
    \includegraphics{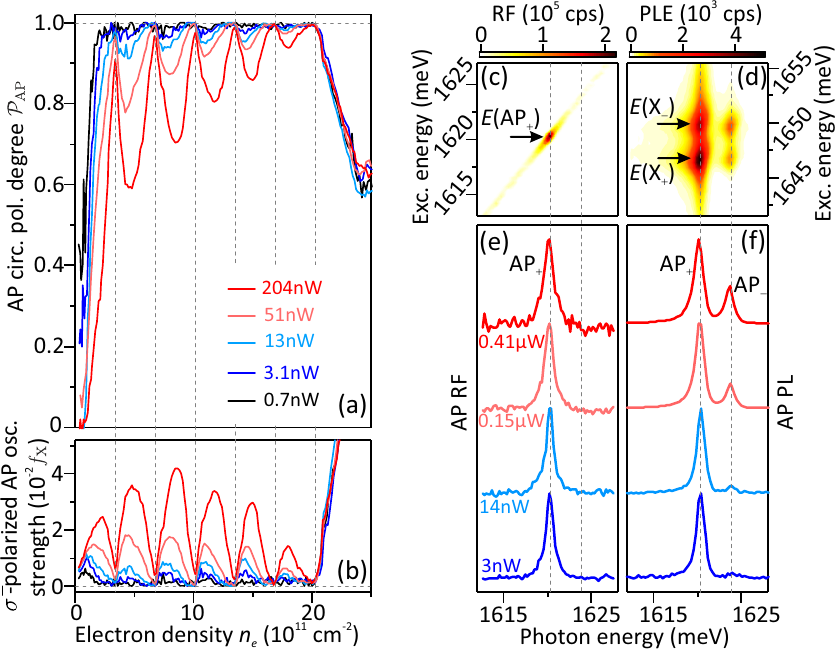}
	\caption{(a,b)~Polarization degree $\mathcal{P}_\mathrm{AP}$ of the AP transition~(a) and the oscillator strength $f_\mathrm{AP_-}$ of the $\sigma^-$-polarized AP resonance~(b) determined as a function of the electron density $n_e$ based on the reflectance contrast measurements carried out at $T=80$~mK, $B=14$~T and for different (indicated) powers of the white-light excitation [the spectrum of exploited white-light source was the same as in Fig.~\ref{fig:Fig1}(c)]. Vertical dashed lines mark the densities corresponding to subsequent integer filling factors. (c,d)~Color-scale plots displaying the AP resonant fluorescence~(c) and AP photoluminescence excited~(d) using a tunable, linearly-polarized, spectrally-narrow laser with power of $0.15\ \mu$W. The spectra were detected in orthogonal linear polarization, for fixed $n_e$ corresponding to $\nu\approx0.8$, and at the same $B$ and $T$ as in panels~(a,b). In both plots the dark counts of the CCD camera were subtracted. The RF data were additionally corrected for background signal stemming from imperfectly suppressed laser by subtracting the reference RF signal measured at charge-neutrality. (e,f)~The AP RF~(e) and PL spectra~(f; quasi-resonantly-excited via the higher-energy X$_-$ resonance and averaged over 2-meV-wide excitation energy window) obtained for different powers of the tunable laser. For clarity, the spectra are vertically offset and normalized.\label{fig:Fig2}}
\end{figure}

This unusual temperature-dependence suggests that the observed valley depolarization does not occur in the ground state of the electron system. To verify this claim, we repeat the polarization-resolved reflectance contrast measurements at $T=80$~mK for different powers of the white-light excitation. As shown in Figs~\ref{fig:Fig2}(a,b), the amplitude of $\nu$-dependent oscillations of $\mathcal{P}_\mathrm{AP}$ and $f_\mathrm{AP_-}$ markedly increases for larger powers. Concurrently, the oscillations become indiscernible for powers lower than a few~nW, but the initial drop of $\mathcal{P}_\mathrm{AP}$ at low $\nu<1$ remains pronounced even below 1~nW. Given that all of the utilized excitation powers are significantly lower than the cooling power of our dilution unit (of a few~$\mu$W~\cite{Kroner_PRA_2014}), these observations indicate that the loss of electron spin-valley polarization arises due to exciton-mediated spin-valley-flip of electrons.

In order to support this conclusion and further exclude any heating-related origin of the investigated effect, we perform RF measurements of the AP resonance using a spectrally-narrow tunable laser. In these experiments the gate voltage is fixed at a value corresponding to $\nu\approx0.8$. Moreover, the reflected light is collected in cross-linear-polarization with respect to the laser, which enables us to suppress the laser background and to address the AP$_+$ and AP$_-$ transitions with equal probabilities. In such a resonant scheme, each of the AP$_\pm$ resonances may be exited only if there are electrons residing in $K^\mp$ valley in the absence of the excitons. Fig.~\ref{fig:Fig2}(c) displays an example RF spectrum acquired under such conditions. It features only one, lower-energy AP$_+$ resonance, which evidences complete polarization of the electrons in their ground state. This finding remains valid independently of the utilized laser power [Fig.~\ref{fig:Fig2}(e)], including the powers for which the electrons are already sizably depolarized under broadband white-light excitation [cf.~Fig.~\ref{fig:Fig2}(a)]. Furthermore, the electronic depolarization is also induced by the resonant laser when its energy is tuned to either of the two Zeeman-split exciton states (X$_\pm$). This is revealed by Figs~\ref{fig:Fig2}(d,f) presenting the corresponding AP PL spectra obtained under such excitation conditions, where we observe both AP$_\pm$ peaks with their intensity ratio increasing with the laser power.

\begin{SCfigure*}
    \includegraphics{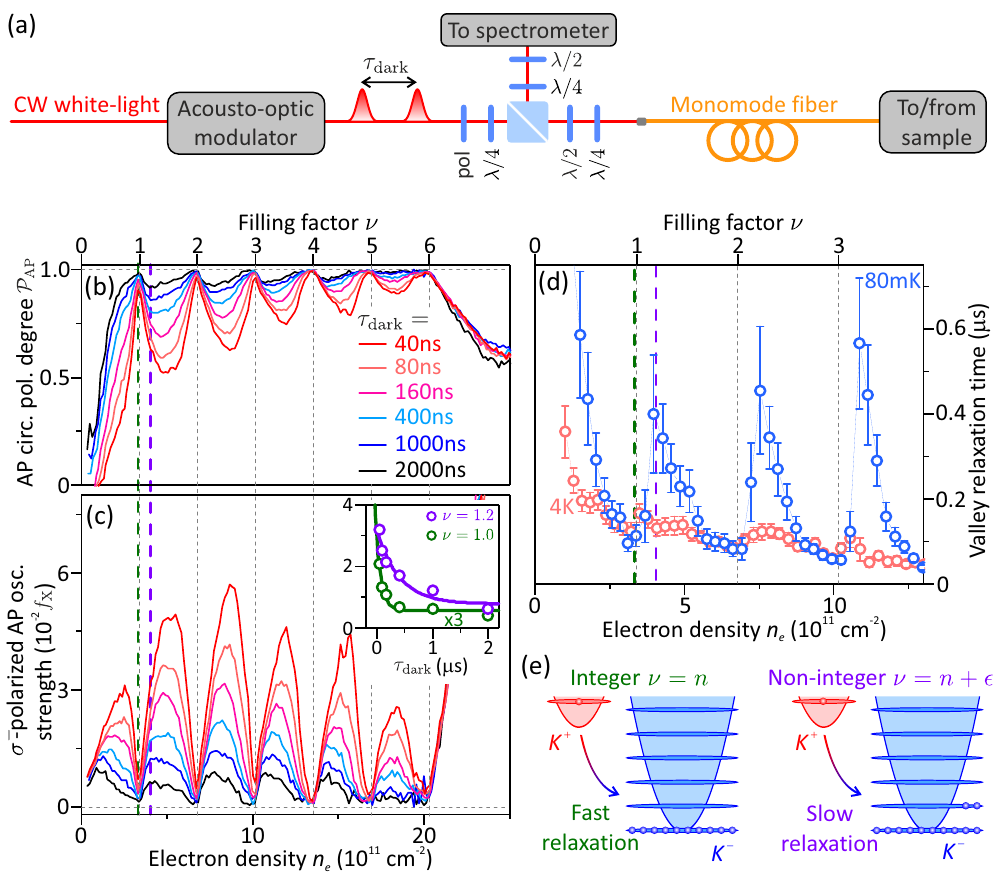}
	\caption{(a)~Simplified illustration of the experimental setup used in the time-resolved measurements of the electron relaxation dynamics. The sample is illuminated with a train of $\sim$15-ns-long, Gaussian pulses obtained from a CW white-light beam using an acousto-optic modulator. The time-integrated and polarization-resolved reflection contrast spectra are acquired as a function of pulse separation $\tau_\mathrm{dark}$ for a fixed pulse energy. (b,c)~Electron-density dependence of the AP polarization degree $\mathcal{P}_\mathrm{AP}$~(b) and the oscillator strength $f_\mathrm{AP_-}$ of the $\sigma^-$-polarized AP resonance~(c) obtained for different values of $\tau_\mathrm{dark}$ at $B=14$~T and $T=80$~mK. The inset in panel~(c) displays $f_\mathrm{AP_-}$ determined for two selected $\nu=1.0$ and $1.2$ versus $\tau_\mathrm{dark}$ (note that the former dataset was multiplied by 3). The solid lines represent the fits of the data points with the exponential decay profiles $\propto\exp(-\tau_\mathrm{dark}/\tau_\mathrm{relax})+\mathrm{const.}$ allowing to extract the electron valley relaxation time $\tau_\mathrm{relax}$. (d)~The value of $\tau_\mathrm{relax}$ determined as a function of the electron density at $B=14$~T for two different temperatures $T=80$~mK and $T=4$~K (the data were binned in $\Delta\nu\approx0.1$ increments). (e)~Cartoons schematically illustrating the filling factors at which the electron valley relaxation is faster and slower.\label{fig:Fig3}}
\end{SCfigure*}

The above results unequivocally demonstrate that---otherwise fully spin-polarized---itinerant electrons undergo spin-valley-flips in the presence of optically injected excitons. After each spin-flip event, an electron remains in the $K^+$ valley until it relaxes back to the $K^-$ one, which gives rise to a finite electron spin-depolarization probed in our time-integrated studies. The corresponding depolarization degree is naturally expected to increase for low electron densities $n_e$ (when the electrons with flipped spins constitute a larger fraction of the total $n_e$), partially explaining why $\mathcal{P}_\mathrm{AP}$ exhibits a sharp decrease around low $\nu$. At the same time, the depolarization efficiency is also proportional to both the exciton injection rate (i.e., excitation power) and the valley relaxation time. Since the latter has been previously demonstrated to be longer for the holes (presumably owing to their larger spin-orbit splitting)~\cite{Dey_PRL_2017,Goryca_SciAdv_2019}, one may expect the depolarization in this case to be more efficient than for the electrons. This conjecture remains in stark contrast with our experimental results [cf.~Fig.~\ref{fig:Fig1}(c)], indicating that it is a difference in the exciton-induced carrier spin-flip rates that is a dominant factor responsible for more prominent electron spin depolarization. We speculate that the larger electron spin-flip probability stems from a very small splitting between the bright and dark intra-valley excitons with the opposite electron spin orientation~\cite{Lu_2D_2019,Robert_NC_2020}: if these states had the same energy, spin-orbit interaction could turn a bright exciton into an intra-valley dark one, and upon spin-preserving valley relaxation of the electron, an inter-valley dark exciton may be formed. Upon subsequent recombination of the hole with a same-valley Fermi-see electron, a net valley-flip excitation would be generated.

Taking advantage of the opportunity to drive the electrons out of equilibrium, we analyze their spin-valley relaxation dynamics. To this end we perform a time-resolved experiment [Fig.~\ref{fig:Fig3}(a)], in which we monitor the steady-state reflectance contrast of the sample excited with a train of equidistant, $\sim$15-ns-long white-light pulses separated by a dark period $\tau_\mathrm{dark}$, which are produced by an acousto-optic modulator. Figs~\ref{fig:Fig3}(b,c) present the $n_e$-evolution of $\mathcal{P}_\mathrm{AP}$ and $f_\mathrm{AP_-}$ obtained at $T=80$~mK for a fixed pulse intensity but different $\tau_\mathrm{dark}$. As expected, the $\nu$-dependent oscillations of both quantities become less prominent for longer $\tau_\mathrm{dark}$. Interestingly, however, for such $\tau_\mathrm{dark}$ the profile of $f_\mathrm{AP_-}$ evolution between subsequent integer $\nu=n$ and $n+1$ acquires clearly asymmetric shape: while at short $\tau_\mathrm{dark}$ the local maximum of $f_\mathrm{AP_-}$ occurs around half-integer $\nu\approx n+1/2$, for longer $\tau_\mathrm{dark}$ the $f_\mathrm{AP_-}$ exhibits a sharp rise followed by a smoother decrease, resulting in a shift of the local maximum towards $\nu=n$. This observation implies that the dependence of $f_\mathrm{AP_-}$ on $\tau_\mathrm{dark}$ is distinct for various $\nu$ [as seen in the inset to Fig.~\ref{fig:Fig3}(c)], demonstrating that the electron relaxation time $\tau_\mathrm{relax}$ must change with $\nu$. 

Fig.~\ref{fig:Fig3}(d) displays the $\tau_\mathrm{relax}$ determined by fitting the exponential decay profiles to $f_\mathrm{AP_-}(\tau_\mathrm{dark})$ measured at different $\nu$. The extracted $\tau_\mathrm{relax}$ exhibits prominent oscillations with $\nu$: it is the shortest when the highest-energy $n$th LL is completely filled, then steeply increases by a factor of $\sim5$ (to about 400--600~ns) when the next $(n+1)$th LL starts to be occupied, before it slowly decreases again reaching approximately the initial value when $(n+1)$th LL gets almost full [Fig.~\ref{fig:Fig3}(e)]. These periodic fluctuations of $\tau_\mathrm{relax}$ entail larger (smaller) electron steady-state spin polarization around integer (half-integer) $\nu$, which is the underlying reason for the oscillating electron depolarization degree observed in our 80-mK CW experiments (cf. Figs~\ref{fig:Fig1}~and~\ref{fig:Fig2}). Importantly, the $\nu$-dependent variation of $\tau_\mathrm{relax}$ is found to be almost fully suppressed at $T=4$~K for $\nu \gtrsim 0.5$ [Fig.~\ref{fig:Fig3}(d)], where $\tau_\mathrm{relax}$ exhibits a decreasing tendency with $\nu$ yielding 50--200~ns in the analyzed $\nu<4$ range. The values we find for $\tau_\mathrm{relax}$ in this regime are in agreement with the electron spin relaxation times determined for various TMD monolayers in some of the previous reports~\cite{Song_NL_2016,Dey_PRL_2017,Goryca_SciAdv_2019}.

Owing to spin-valley locking, phonon-mediated relaxation of the excited states in TMD monolayers is strongly suppressed at low temperatures by energy-momentum conservation. Even though we cannot experimentally rule out the influence of coupling to a phonon reservoir on the spin relaxation dynamics, its marked ground-state dependence hints at a central role played by electronic correlations induced by strong Coulomb interactions. Such correlations are suppressed when the electrons form an IQH state at $\nu=n$, but may become pronounced whenever the system gets occupied by excess electrons ($\nu=n+\epsilon$) or excess holes ($\nu=n-\epsilon$)~\cite{koulakov1996charge,fogler1996ground,fogler2002stripe}. We speculate that this gives rise to substantially longer $\tau_\mathrm{relax}$ revealed by our mK-experiments at $\nu=n+\epsilon$ (presumably with an electron/hole asymmetry being responsible for the absence of an analogous effect at $\nu=n-\epsilon$). This speculation is consistent with a similar slow-down of the relaxation dynamics demonstrated previously for WSe$_2$/WS$_2$ hetero-bilayer hosting a Mott-like CI state~\cite{Regan2020}. Our conjecture is further supported by recent experiments~\cite{Smolenski_arXiv_2020} evidencing the formation of the WC ground state in a monolayer system at $B=14$~T for $\nu\lesssim0.5$, which is the $\nu$-range where the prolongation of $\tau_\mathrm{relax}$ is clearly the most prominent. The reported WC melting temperature exceeds 4~K, which may explain why the increase of $\tau_\mathrm{relax}$ is observed in this regime not only at $T=80$~mK, but also at $T=4$~K [cf. Fig.~\ref{fig:Fig3}(d)]. The lack of a similar increase at $\nu=n+\epsilon>1$ for $T=4$~K may be a consequence of a more fragile nature of the corresponding correlated states, which might melt for $T<4$~K.

The strong filling-factor dependence of spin-valley relaxation uncovered by our work paves the way towards future explorations of non-equilibrium dynamics of electrons in atomically thin semiconductors. In parallel, large efficiency of light-induced electron spin depolarization mechanism utilized in our experiments indicates that excitons might not constitute a non-destructive probe of the electronic system in TMD monolayers even when the electron density is orders of magnitude larger than that of the excitons.

\begin{acknowledgments}
We thank P. Back and A. Popert for fabricating the investigated devices. This work was supported by the Swiss National Science Foundation (SNSF) under Grant No. 200021-178909/1. K.W. and T.T. acknowledge support from the Elemental Strategy Initiative conducted by the MEXT, Japan (Grant Number JPMXP0112101001) and JSPS KAKENHI (Grant Numbers 19H05790 and JP20H00354).
\end{acknowledgments}

%\bibliography{Reference}
%

\end{document}

% ----- Title -----

\title{Supplemental Material for \\ ``Spin-valley relaxation dynamics of Landau-quantized electrons in MoSe$_2$ monolayer''}

\author{T. Smole\'nski}
\affiliation{\ETH}

\author{K.~Watanabe}
\affiliation{\NIMSRCFM}

\author{T.~Taniguchi}
\affiliation{\NIMSICMN}

\author{M. Kroner}
\affiliation{\ETH}

\author{A. Imamo\u{g}lu}
\affiliation{\ETH}

\maketitle

\section{Analysis of the exciton-polaron spectral profiles}
\label{sec:TM_fitting}

\begin{figure}[b!]
    \includegraphics{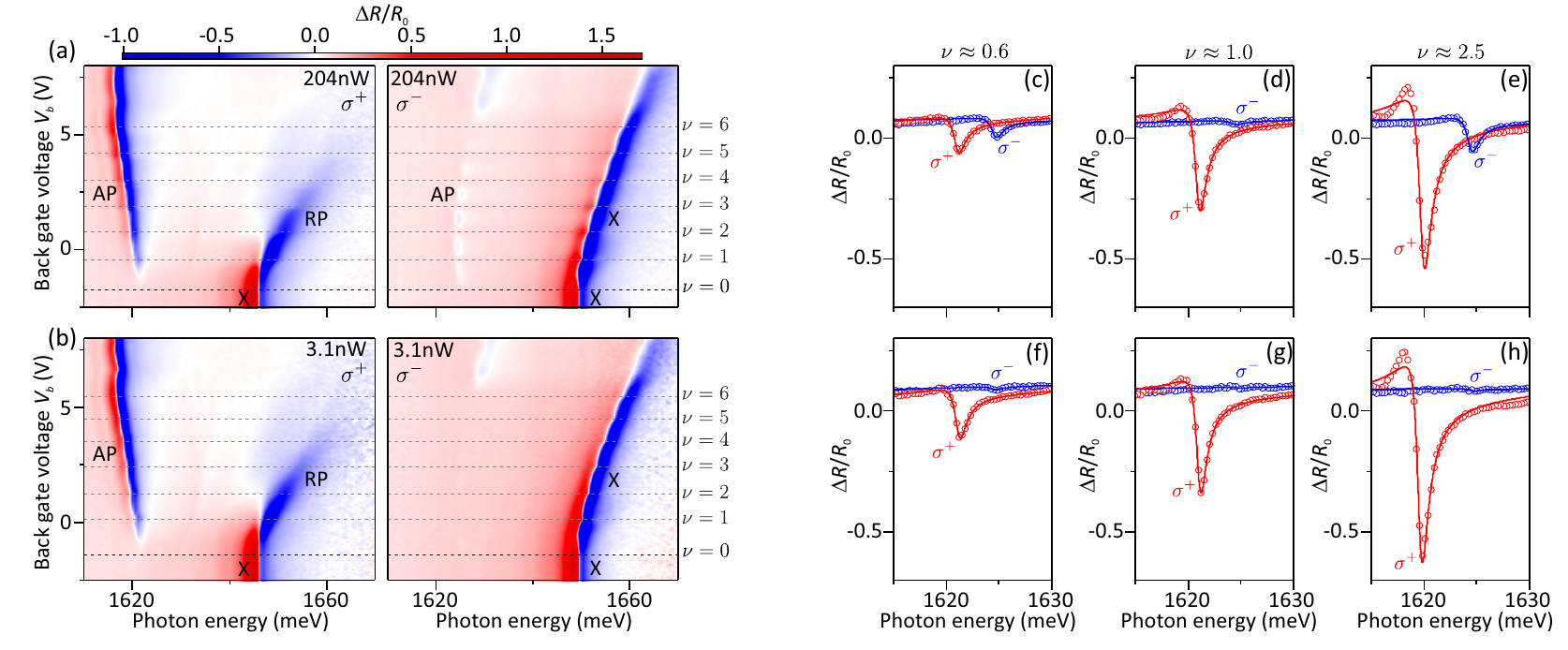}
	\caption{(a,b) Color-scale plots showing back-gate-voltage evolutions of the reflectance contrast spectra measured at $T=80$~mK, $B=14$~T and for two different powers of the white-light excitation: $204$~nW~(a) and 3.1~nW~(b). The data were acquired in the two circular polarizations: $\sigma^+$ (left) and $\sigma^-$ (right). The horizontal dashed lines mark the voltages corresponding to subsequent integer filling factors. To facilitate the analysis of weak exciton-polaron transitions, the spectra were convolved with a discretized Gaussian profile of $\sqrt{2}\sigma$ corresponding to the separation of subsequent CCD pixels (i.e., $\sqrt{2}\sigma\approx0.13$~nm). (c-h) Line cuts of the maps in panels (a,b) displaying $\sigma^\pm$-polarized spectra in the visinity of the attractive polaron resonances acquired at gate voltages corresponding to different filling factors: $\nu\approx0.6$~(c,f), $\nu\approx1.0$~(d,g), and $\nu\approx2.5$~(e,h). The solid lines mark the fits of the experimental data with the transfer matrix approach, based on which we extract the oscillator strengths of AP$_\pm$ transitions.\label{fig:FigS1}}
\end{figure}

In our resonant reflection experiments, we analyze the gate-voltage evolution of the reflectance contrast signal $\Delta R/R_0=(R-R_0)/R_0$ determined based on the white-light reflectance spectra $R$ and $R_0$ measured in the region with and without the MoSe$_2$ flake, respectively. To probe the $\sigma^\pm$-polarized response of the investigated device, we simultaneously adjust the polarizations of exciting and detected light (thus keeping both beams always co-polarized), which enables us to sizably enhance the polarization contrast. 

Due to the interference of light reflected off the MoSe$_2$ monolayer and off the other interfaces in our heterostructure, the excitonic resonances appearing in $\Delta R/R_0$ are not purely Lorentzian, but instead acquire a partially dispersive lineshape. To account for this effect, we describe the measured spectral profiles using a transfer-matrix formalism~\cite{Back_PRL_2018,Scuri_PRL_2018}. In this approach, the thicknesses of the top and bottom hBN layers of are taken, respectively, as $t_t\approx74$~nm and $t_b\approx91$~nm based on the atomic force microscope (AFM) measurements (see Ref.~\cite{Smolenski_arXiv_2020} for further details regarding the structure of the investigated sample), while the hBN refractive index is assumed to be equal to $n_\mathrm{hBN}=2.1$~\cite{Lee_PSSB_2019}. The $\sigma^\pm$-polarized attractive polaron (AP$_\pm$) transitions in the MoSe$_2$ monolayer are described using the following, complex susceptibility $\chi^\pm_\mathrm{AP}(E)=-(\hbar c/E^\pm_\mathrm{AP})\gamma^\pm_\mathrm{rad}/(E-E^\pm_\mathrm{AP}+i\gamma^\pm_\mathrm{nrad}/2)$~\cite{Scuri_PRL_2018}, where $E$ is the photon energy, $E^\pm_\mathrm{AP}$ represents the AP$_\pm$ energy, while $\gamma^\pm_\mathrm{nrad}/\hbar$ and $\gamma^\pm_\mathrm{rad}/\hbar$ correspond, respectively, to non-radiative and a free-space radiative AP$_\pm$ decay rates (the latter being directly proportional to the oscillator strength $f_{\mathrm{AP}_\pm}$). Under these assumptions, we compute the reflectance spectra $R$ and $R_0$ on and off the MoSe$_2$ monolayer in the spectral vicinity of the AP$_\pm$ resonances and thus obtain $\Delta R/R_0$. For a quantitative comparison with the experimental data, this quantity is corrected for the residual, smooth background (e.g., originating from the proximity of the exciton/repulsive polaron transition, which is not included in the susceptibility $\chi^\pm_\mathrm{AP}$) by adding a constant term $C$ and, in case of the power-dependent data from Figs~2~and~\ref{fig:FigS1}, also a phenomenological, linearly increasing background $s(E-E^\pm_\mathrm{AP})$ with a constrained positive slope $0<s<10\ \mu\mathrm{m}^{-1}$. 

To extract the electron-density-dependent oscillator strengths of the AP$_\pm$ resonances, the above model is fitted to the reflectance contrast spectra. In order to facilitate the analysis of weak optical transitions, some of the measured spectra (corresponding to the data shown in Figs~2,~3,~\ref{fig:FigS1},~and~\ref{fig:FigS2}) are first convolved with a discretized Gaussian profile of $\sqrt{2}\sigma\approx0.13$~nm corresponding to the separation of subsequent pixels in our CCD camera. The fitting is always carried out in a narrow wavelength range around the analyzed resonance, with the width (of a few~nm) being adjusted separately for each spectrum based on the extracted resonance linewidth. In case of the $\sigma^+$-polarized spectra featuring stronger attractive polaron transition, all four variables $E^+_\mathrm{AP}$, $\gamma^+_\mathrm{nrad}$, $\gamma^+_\mathrm{rad}$, $C$ (and $s$ in case of the data in Figs~2~and~\ref{fig:FigS1}) are treated as independent fitting parameters except for the low-electron-density range, where the AP$_+$ transition becomes weak (at $\nu<\nu_c$ with $\nu_c$ between $0.2-0.8$ depending on the dataset). For the spectra acquired at such $n_e$, the non-radiative broadening $\gamma^+_\mathrm{nrad}$ (mainly controlling the resonance linewidth) is fixed at a value obtained based on the analysis of one or a few spectra taken at subsequent voltages falling just beyond the invoked range $\nu<\nu_c$. Similar procedure is applied for fitting $\sigma^-$-polarized spectra acquired at low $n_e$. In case of this circular polarization, the value of $\gamma^-_\mathrm{nrad}$ is also fixed when fitting the spectra taken around some integer filling factors $\nu=n\le6$ (where the AP$_-$ peak is barely discernable) at an average value of $\gamma^-_\mathrm{nrad}$ determined for $\nu$ lower and higher than $n$ at which the AP$_-$ transition becomes resolvable again. Finally, during the fitting of $\sigma^-$-polarized spectra acquired for lower excitation powers (in case of the data in Figs~2~and~\ref{fig:FigS1}) or longer dark periods (in case of the data in Fig.~3) at $\nu\lesssim6$, the $\gamma^-_\mathrm{nrad}(n_e)$ is fixed at the value obtained at the same $n_e$ for the highest excitation power or the shortest dark period, respectively (with possible linear interpolation between subsequent $\gamma^-_\mathrm{nrad}(n_e)$ values in case the grids of $n_e$ values differ between various datasets). 

The above fitting procedure enables us to precisely reproduce the lineshapes of the AP resonances measured in two polarizations for different gate voltages, as seen in Fig.~\ref{fig:FigS1} showing example fits of $\Delta R/R_0$ spectra acquired at two different white-light excitation powers. On this basis we extract the electron-density-dependent $\gamma^\pm_\mathrm{rad}$ for both AP$_\pm$ transitions (note that in case of the time-resolved data at $T=4$~K, such $\gamma^\pm_\mathrm{rad}(n_e)$ values for each dark period are averaged over a few separate measurements in order to reduce the noise, so are also the values of $\gamma^-_\mathrm{nrad}(n_e)$ obtained based on the shortest-dark-period dataset and used for fitting the spectra at longer $\tau_\mathrm{dark}$). To convert $\gamma^\pm_\mathrm{rad}$ values into oscillator strengths, we also extract the free-space radiative decay rate $\gamma^\mathrm{X}_\mathrm{rad}/\hbar$ of the neutral exciton that is assumed to be independent of the excitation conditions (within the ranges explored in this study). To this end, we perform an analogous transfer-matrix fit of the exciton resonance in the charge neutrality. The obtained value of $\gamma^\mathrm{X}_\mathrm{rad}$ is then averaged over the two circular polarizations and over a gate voltage range at which the MoSe$_2$ is devoid of itinerant carriers. This enables us to express the $n_e$-dependent oscillator strengths $f_{\mathrm{AP}_\pm}(n_e)$ of AP$_\pm$ resonances in the units of the exciton oscillator strength $f_\mathrm{X}$ as $f_{\mathrm{AP}_\pm}(n_e)/f_\mathrm{X}=\gamma^\pm_\mathrm{rad}(n_e)/\gamma^\mathrm{X}_\mathrm{rad}$, which is how we determine the $f_{\mathrm{AP}_\pm}(n_e)$ plotted in all figures in the main text.

\section{Calibration of the electron density}
\label{sec:ne_calibration}

\begin{SCfigure*}
    \includegraphics{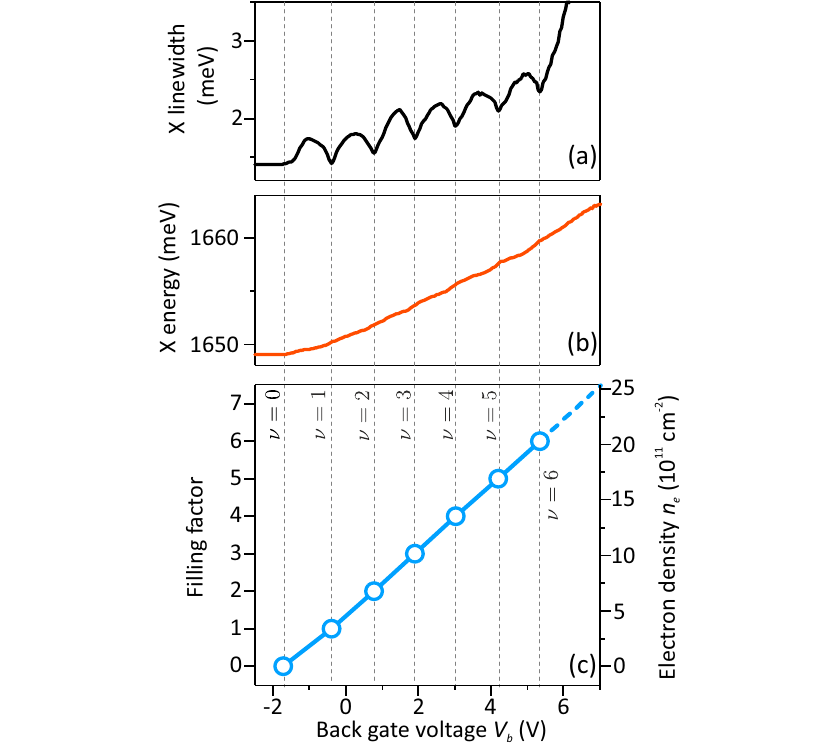}
	\caption{(a,b)~Back-gate-voltage evolution of the linewidth~(a) and energy~(b) of the $\sigma^-$-polarized neutral exciton transition determined by fitting its spectral profile with dispersive Lorentzian formula (given in Refs~\cite{Smolenski_arXiv_2020,Smolenski2019}) in the reflectance contrast spectra from Fig.~\ref{fig:FigS1} (measured at $T=80$~mK, $B=14$~T, and 204~nW white-light excitation power). The vertical dashed lines mark the voltages corresponding to integer filling factors. (c) Gate voltages corresponding to subsequent integer filling factors extracted by tracing the positions of the linewidth minima ($1\le\nu\le6$) or by determining the voltage, at which the exciton begins to blueshift ($\nu=0$). The solid line represents the linear interpolation between the subsequent data points that is used to calibrate the electron density for $0\le\nu\le6$. The dashed line depicts the linear extrapolation of the slope between the points at $\nu=5$ and $\nu=6$ that is used to calibrate $n_e$ for $\nu>6$.\label{fig:FigS2}}
\end{SCfigure*}

In the experiments reported in the main text, the electron density $n_e$ is controlled by applying a voltage $V_g$ to either top or back gate electrodes. In case the datasets from Fig.~1, the $n_e$ is tuned by ramping the top gate, while keeping the TMD monolayer and the back gate grounded. When acquiring the data in Figs~2~and~3 (as well as in Figs~\ref{fig:FigS1}~and~\ref{fig:FigS2}), the electrons are injected to (still grounded) TMD monolayer with the back gate, while keeping the other gate electrically floating. In order to precisely calibrate the value of $n_e$ in both of these gating schemes, we analyze the Shubnikov--de Haas oscillations of the exciton linewidth $\gamma^-_\mathrm{X}$ in the $\sigma^-$ polarization, which is extracted from the measured reflectance spectra by fitting the exciton spectral profile with a phenomenological, dispersive Lorentzian formula (as detailed in Ref.~\cite{Smolenski_arXiv_2020}). As recently demonstrated in our previous study~\cite{Smolenski2019}, such linewidth exhibits a prominent minima when the system enters an integer quantum Hall state, as seen in the example dataset shown in Fig.~\ref{fig:FigS2}(a) (and also in Fig.~1(d) in the main text displaying an analogous dataset obtained from a separate measurement). By tracing the positions of these minima we extract the gate voltages $V_{g,n}$ corresponding to subsequent integer filling factors $\nu=n$ and hence to the electron densities of $neB/h$ (where $B=14$~T). This procedure can only be applied for $1\le\nu=n\le6$, since for higher $\nu$ the linewidth oscillations are no longer resolvable due to the excessive broadening of the exciton transition that occurs when the Fermi level exceeds the valley-Zeeman splitting of the conduction band. In parallel, we also determine the voltage $V_{g,0}$ corresponding to the onset of filling the MoSe$_2$ with electrons (i.e., $\nu=0$) that is extracted as a value of $V_g$ at which the exciton transition begins to blueshift [cf.~Fig.~\ref{fig:FigS2}(b)]. The electron density $n_e$ at a given $V_g$ (for $\nu\le6$) is then determined by linear interpolation between subsequent integer filling factors as $n_e(V_g)=(eB/h)\cdot[n+(V_g-V_{g,n})/(V_{g,n+1}-V_{g,n})]$ for $V_{g,n}<V_g<V_{g,n+1}$ and $n<6$, as marked by the solid line in Fig.~\ref{fig:FigS2}(c). In case of $\nu>6$, the $n_e$ is in turn obtained by linear extrapolation of the slope between $\nu=5$ and $\nu=6$, i.e., $n_e(V_g)=(eB/h)\cdot[6+(V_g-V_{g,6})/(V_{g,6}-V_{g,5})]$, as marked by the dashed line in Fig.~\ref{fig:FigS2}(c).

The above procedure is used to calibrate the electron density in all the figures in the main text. Note that to account for optically-induced changes in the electron density, such calibration is carried out separately for every dataset acquired at different excitation conditions (i.e., at different white-light power or for different duration of the dark period in case of time-resolved experiments).

\section{Complementary datasets obtained for a second device}
\label{sec:second_dev}

\begin{SCfigure*}[][b]
    \includegraphics{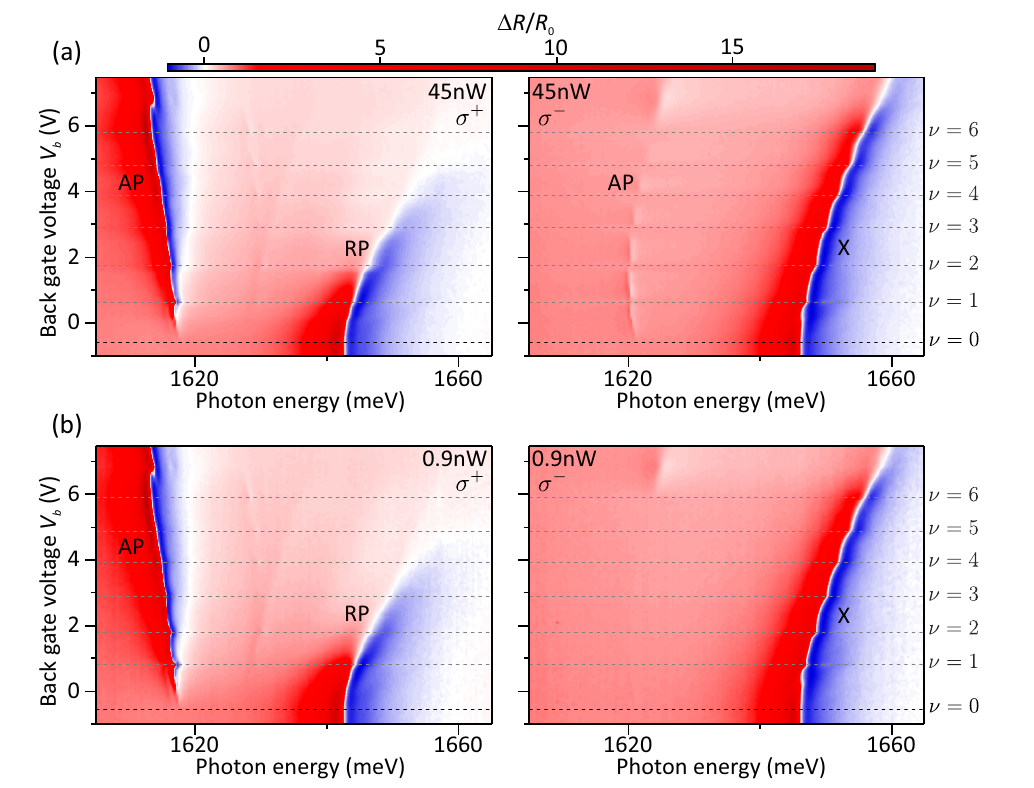}
	\caption{(a,b)~Color-scale maps showing back-gate-voltage evolution of the reflectance contrast spectra acquired for the second device in the two circular polarizations: $\sigma^+$ (left) and $\sigma^-$ (right). The spectra were measured at $T=80$~mK, $B=14$~T, and for two different powers of the white-light excitation: $45$~nW~(a) and $0.9$~nW~(b). During the measurements the top gate voltage was fixed at $V_t=1$~V.\label{fig:FigS3}}
\end{SCfigure*}

As stated in the main text, analogous $\nu$-dependent oscillations of the electron spin-valley polarization were also revealed by our time-integrated experiments carried out on a second device. The investigated spatial region of this device featured a similar layer structure to that of the main device, consisting of a dual-gated MoSe$_2$ monolayer that was electrically contacted with a FLG flake and fully encapsulated between top and bottom hBN flakes. Fig.~\ref{fig:FigS3} displays a back-gate-voltage evolution of the $\sigma^\pm$-polarized reflectance contrast spectra acquired for this device at $T=80$~mK, $B=14$~T, and for two different powers of white-light excitation of 45~nW and 0.9~nW. In the dataset obtained for the higher power, [Fig.~\ref{fig:FigS3}(a)] we clearly observe a $\nu$-dependent oscillations of the intensity of the AP$_-$ resonance in $\sigma^-$ polarization: the resonance becomes stronger around half-integer filling factors and gets weaker close to integer ones. Crucially, these oscillations disappear upon lowering the excitation power to 0.9~nW, when the AP$_-$ transition becomes almost indiscernible [Fig.~\ref{fig:FigS3}(b)]. These findings are fully consistent with our observations for the main device, which further supports the conclusions drawn in the main text.